# Sustainable Carbon-Aware and Water-Efficient LLM Scheduling in Geo-Distributed Cloud Datacenters

Hayden Moore, Sirui Qi
*Colorado State University*
Fort Collins, USA
{hayden.moore, alex.qi}@colostate.edu

Ninad Hogade, Dejan Milojicic, Cullen Bash
*Hewlett Packard Labs*
Milpitas, USA
{ninad.hogade, dejan.milojicic, cullen.bash}@hpe.com

Sudeep Pasricha
*Colorado State University*
Fort Collins, USA
sudeep@colostate.edu

## ABSTRACT

In recent years, Large Language Models (LLM) such as ChatGPT, CoPilot, and Gemini have been widely adopted in different areas. As the use of LLMs continues to grow, many efforts have focused on reducing the massive training overheads of these models. But it is the environmental impact of handling user requests to LLMs that is increasingly becoming a concern. Recent studies estimate that the costs of operating LLMs in their inference phase can exceed training costs by 25× per year. As LLMs are queried incessantly, the cumulative carbon footprint for the operational phase has been shown to far exceed the footprint during the training phase. Further, estimates indicate that 500 ml of fresh water is expended for every 20–50 requests to LLMs during inference. To address these important sustainability issues with LLMs, we propose a novel framework called SLIT to co-optimize LLM quality of service (time-to-first token), carbon emissions, water usage, and energy costs. The framework utilizes a machine learning (ML) based metaheuristic to enhance the sustainability of LLM hosting across geo-distributed cloud datacenters. Such a framework will become increasingly vital as LLMs proliferate.

## KEYWORDS
Large language model, Carbon emissions, Water, Energy cost

## 1 Introduction

Large Language Models (LLMs) are starting to dominate cloud computing workloads, with adoption across the domains of e-commerce, finance, healthcare, and search engines. Their growth has been led by OpenAI's ChatGPT and other generative artificial intelligence (AI) tools with this service alone reaching a monthly active user base of close to 100 million users by 2022 [1], which has almost doubled in the past few years [2]. This growth in usage has led to a proportionally rapid increase in the associated energy consumption of cloud servers hosting LLMs, with current estimates of 25-33% compound annual growth rate [3].

The increase in LLM-driven energy usage in cloud datacenters is a major concern, as prior to the introduction of LLMs, datacenters were already estimated to consume approximately 1-2% of global electricity [4]. LLMs are threatening to increase this proportion significantly. While LLM training is often considered the culprit, surprisingly it is LLM inference that dominates costs. LLM inference consumes 25× more computational resources per year than the initial training [5]. Another aspect that cannot be ignored is the carbon emission footprint of LLMs. It is estimated that today's cloud computing infrastructure emits approximately 0.6% of global carbon emissions [5]. For large-scale deployments like Google Search, LLM inference could have 1,400× more carbon footprint than training annually [1]. Thus, carbon emissions of datacenters used for LLMs represent a key climate challenge.

Yet another aspect of the environmental footprint of LLMs is their water usage in cloud datacenters. Water is becoming an increasingly scarce resource within urban centers and it is currently estimated that 30% of all cities with a population over a million suffer from yearly water scarcity [6]. Cloud datacenters already consume a large amount of water, with a medium size datacenter (15 MW) consuming the same amount of water as two 18-hole golf courses [7]. LLM inference is exacerbating water use in these datacenters, which can no longer be ignored.

In summary, there is a critical need for sustainable LLM hosting, particularly for the LLM inference phase. Practically, we are referring to the minimization of carbon emissions, water use, and energy costs, while ensuring acceptable performance (measured via the metric time-to-first token (TTFT), which determines when the first response can be sent back to the user) during LLM inference. In this paper, we propose a novel multi-objective optimization framework to make scheduling decisions for LLM requests while optimizing for the TTFT, carbon emissions, water usage, and energy costs of each LLM inference task. Our proposed framework for sustainable geo-distributed LLM request scheduling called *SLIT* utilizes a metaheuristic that combines machine learning (ML) and an evolutionary algorithm (EA) to generate a Pareto optimal solution set in real-time, to promote sustainable LLM inference across application domains.

The novel contributions of our work are:

- We formulate a multi-objective LLM inference scheduling problem that co-optimizes the TTFT, carbon emissions, water usage, and energy cost of each LLM request.
- We propose a new framework called SLIT that utilizes a metaheuristic combining ML and EA algorithms.
- We model the carbon footprint, water usage, and energy costs across geo-distributed LLM-hosting cloud datacenters consisting of heterogeneous hardware.
- We compare SLIT with several state-of-the-art LLM inference scheduling frameworks and show that our framework can provide a better set of holistic solutions.

The rest of the paper is organized as follows. We discuss related work in Section 2. Section 3 discusses the models utilized in this work. Section 4 describes the problem formulation. Our proposed framework is discussed in Section 5. The results of our framework and comparison works are presented in Section 6. Lastly, we present concluding remarks in Section 7.

## 2 Related Work

Scheduling workloads within cloud datacenters has been an extensively studied problem. In the management of cloud datacenters, different metrics are prioritized, with performance and energy being the most popular ones. Performance can be measured in terms of the latency of the workload at the datacenter. Ajmal et al. [8] proposed a hybrid genetic algorithm to reduce the workload latency within datacenters. Hogade et al. [9] used a game theoretic approach that factors in the cost of data transfers to reduce the amount of energy used in datacenters. A survey of the use of ML within cloud datacenter scheduling and resource management was conducted in [10] and discusses many efforts to optimize performance and energy use in cloud datacenters.

As climate change continues to be an increasingly prevalent problem, several solutions have been developed in recent years to reduce the impact of cloud datacenters on the environment. A score-based heuristic was proposed by Chadha et al. [11] which scores the carbon emissions associated with running a workload on a datacenter and demonstrates a reduction in carbon emissions per workload. Qi, et al. [12] proposed a dual optimizer that improves both the throughput and carbon emissions of a cloud workload. Some recent efforts go beyond carbon emissions and also consider water emissions. In [13] a heuristic was proposed to co-optimize performance and water usage for cloud workloads. In [14] a hybrid cloud scheduling heuristic was proposed targeting a more complex geo-distributed network of datacenters.

With the rise of LLMs, new workload management strategies are needed to handle the unique aspects associated with managing the serving of LLMs. One approach proposed by Ryabinin et al. [15] utilized an adaptive rebalancing approach that balances training across multiple heterogeneous devices. However, this scheduling framework cannot be easily applied to an inference workload due to the short and tighter deadlines of LLM inference tasks. A mixed integer linear programming algorithm for LLM inference scheduling was proposed by Mei, et al. [16] which demonstrates effective balancing of requests across heterogeneous hardware. An approach focused on splitting each request into two phases was proposed by Patel et al. [17] which improved throughput in terms of tokens per second. An approach to reduce energy costs and carbon emissions of AI workloads was proposed by in [18], however they did not consider LLMs.

The lack of a sustainability-focused LLM inference scheduling framework for large-scale deployments across geo-distributed cloud datacenters strongly motivates our current work. Unlike any prior work, our framework co-optimizes carbon emissions, water usage, energy costs, and TTFT performance for LLM inference workloads, using a metaheuristic guided approach. It generates a Pareto front of solutions in real-time, to allow a datacenter manager to weigh solutions across sustainability, performance, and energy cost dimensions, and systematically select the best solution as per their unique requirements.

## 3 Modeling

Our proposed framework is designed to service LLM inference requests across geo-distributed datacenters. Each request is mapped to a server node within one of the datacenters on the geo-distributed network. To develop a meaningful scheduling framework, we model datacenters comprehensively in terms of their energy costs, carbon emissions, and water usage for each mapped inference request. Our LLM workload, datacenter, energy, carbon, and water models are discussed in the rest of this section.

### 3.1 LLM Workload Model

We utilize a real LLM trace [19] that is based on requests to ChatGPT over the course of two weeks. Fig. 1 shows the number of LLM token requests over close to 6000 epochs, where each epoch has a length of 15 minutes. From [19] and the figure, we observe two trends: 1) most of the usage is dominated by smaller and older models, and 2) the request intensity changes rapidly.

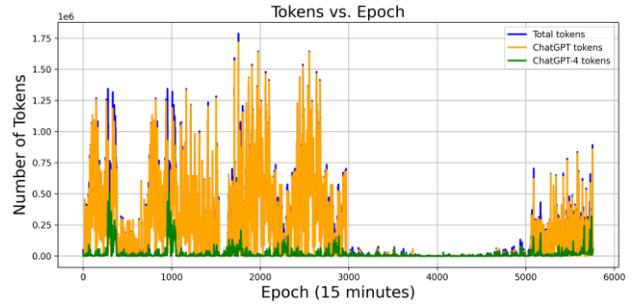

**Fig. 1: The number of LLM tokens requested in each epoch (of 15 minutes) over a two-week period [19].**

We developed a synthetic workload based on the observed trends from [19]. This synthetic workload maps the requests onto two LLM models (Llama-7B and Llama-70B). This allows us to flexibly scale and explore the impact of workload variations across various scheduling frameworks and conditions. LLM requests in the workload are characterized by stringent memory and latency demands, unlike most other cloud workloads.

Two factors contribute to the memory footprint $M_i$ of an LLM inference request $i$. The first is the model parameter memory $M_O$ shared across all requests using the same LLM model $O$. The second is the key-value (KV) cache of each request. The memory associated with the KV cache will grow per output token $n$ up to the total of all output tokens $N_i$. The growth of this KV cache is dependent on the size of the underlying model $M_{KV_O,i}$. We can estimate the memory footprint $M_i$ of LLM inference request $i$ as:

$$M_i = N_i \times M_{KV_O,i} + M_O \qquad (1)$$

The overall latency $F_i$ (defined as TTFT) of any singular inference request $i$ must factor in the loading (orchestration) overhead $F_{load,O}$ of the model $O$, the cross-datacenter migration latency $F_{migrate,l_s,l_d}$ (if migration occurs), and the processing time $F_{process,i}$. The loading overhead $F_{load,O}$ is determined by the size of the LLM model and the bandwidth $BW_g$ of the node as:

$$F_{load,O} = M_O/BW_g \qquad (2)$$

Note that when the cumulative memory footprint $M_i$ exceeds the GPU memory capacity of $M_{cap,g}$ of the GPU compute node $g$ that is servicing the LLM request, then additional loading overhead is added to factor in the latency for reassigning and loading the model on a different available node.

A request migration introduces a latency $F_{migrate,l_s,l_d}$ that is determined by the network conditions between the source datacenter $l_s$ and the destination datacenter $l_d$. The migration latency based on inter-datacenter networks is dependent on the number of hops $R_{l_s,l_d}$ between the source and destination datacenters and the latency between two network routers $K_{media}$ [20]. Transmission media is a major factor that determines inter-router latency $K_{media}$. The migration latency can be expressed as:

$$F_{migrate,l_s,l_d} = R_{l_s,l_d} \times K_{media} \qquad (3)$$

The processing time $F_{process,i}$ is how long it takes to process the first token that is returned. This is assumed to be linear to the number of tokens to process and the total execution time $T_{exec,i}$ for each request $i$. The $F_i$ is then all the individual latencies and initial processing time $F_{process,i}$ (= $T_{exec,i}/N_i$) combined. The migration latency is doubled as the input tokens are sent to the destination and the output tokens are sent back to the source:

$$F_i = F_{load,O} + 2 \times F_{migrate,l_s,l_d} + T_{exec,i}/N_i \qquad (4)$$

### 3.2 Datacenter Model

Each datacenter consists of a total of $G_l$ server nodes at each location $l$. These nodes are part of racks set up in a standard hot/cold aisle arrangement and are cooled with a mechanical cooling system [21]. We assume that server nodes are heterogeneous and can have different combinations and amounts of processing capabilities. Each server node $g$ consists of multiple GPUs that all pool their memory during operation.

### 3.3 Energy Cost Model

Energy usage is tracked on a per-node basis with each node $g$ having three associated power states ($pstate$ of $ON, IDLE, OFF$) with each state being a proportion $PR_{pstate}$ of the specified thermal design power $TDP_g$ of the node [22]. The energy usage of each node $E_{IT,g}$ over a time epoch $t$ is dependent on which power state that specific node operates in:

$$E_{IT,g,t} = PR_{pstate} \times TDP_g \times t, pstate \in [ON, IDLE, OFF] \qquad (5)$$

To estimate the information technology (IT) device energy usage $E_{IT,l,t}$ across all nodes $G_l$ at datacenter location $l$, the individual node energy usage amounts must be combined:

$$E_{IT,l,t} = \sum_g^{G_l} E_{IT,g,t} \qquad (6)$$

The mechanical cooling system energy usage is defined as $E_{cooling,l}$. The largest source of $E_{cooling,l}$ is from the computer room air conditioning (CRAC). The energy consumption $E_{CRAC,l}, t$ of the CRAC in epoch $t$ is determined by associated $E_{IT,l,t}$ and the efficiency of the cooling system (defined as $CoP_l$):

$$E_{CRAC,l,t} = E_{IT,l,t}/CoP_l \qquad (7)$$

The chillers and the other hardware found within the mechanical cooling system also consume energy that is roughly equivalent to the energy used by CRAC units $E_{CRAC,l,t}$ [23]. Therefore, the total cooling energy $E_{cooling,l,t}$ can be estimated as:

$$E_{cooling,l,t} = 3 \times E_{CRAC,l,t} \qquad (8)$$

The amount of energy consumed by the supporting internal power conditioning system (e.g., power distribution units, power supply units) $E_{sup,l,t}$ in epoch $t$ is equivalent to about 13% of the energy of the nodes in the datacenter according to [24]:

$$E_{sup,l,t} = 0.13 \times E_{IT,l,t} \qquad (9)$$

Ultimately, the total energy usage $E_{tot,l}$ of any datacenter $l$ during a time epoch $t$ can be calculated by combining all the sources discussed in this section:

$$E_{tot,l,t} = E_{IT,l,t} + E_{cooling,l,t} + E_{sup,l,t} \qquad (10)$$

The cost of energy is reliant on how the utility company charges for usage on the electricity grid. One important control tool that is used by utility companies globally is a time-of-use price $TOU_{l,t}$ which allows for different pricing around peak hours of usage, and factors in the costs due to the specific generation approach (e.g., coal, hydroelectric, solar). The cost $P_{l,t}$ of energy usage at a datacenter location $l$ during time epoch $t$ is given as:

$$P_{tot,t} = \sum_l^L (E_{tot,l,t} \times TOU_{l,t}) \qquad (11)$$

### 3.3 Water Model

Water usage in a datacenter occurs due to cooling on-site and from the generation of electricity off-site. The sources consist of the evaporative $W_{E,l}$, blowdown $W_{B,l}$, and grid $W_{Grid,l}$ water consumptions for a datacenter $l$. The evaporative source of water usage occurs when heated water evaporates through the datacenter cooling towers. To calculate the water lost to evaporation in an epoch $t$, the total heat generated by the IT equipment $H_{IT,l,t}$ and heat capacity of water $H_{water}$ are factored:

$$W_{E,l,t} = H_{IT,l,t}/H_{water} \qquad (12)$$

The use of mechanical cooling involves the use of a coolant which is typically water. The water required for this cooling is potable, whereas the outgoing water must be processed at a wastewater facility [4]. This is due to the buildup of solid chemicals represented as a ratio $D$ within the water. This blowdown water in epoch $t$ can be characterized by:

$$W_{B,l,t} = W_{E,l,t}/(1-D) \qquad (13)$$

Each source of electricity also has an underlying water intensity $WI_l$. Therefore, the use of electricity must also be considered for its water usage, based on the overall water intensity of the grid for datacenter $l$. The water intensity varies dramatically between energy sources, such as wind energy using 0.2L/kWh, and hydropower using 67L/kWh [25]:

$$W_{Grid,l,t} = E_{tot,l,t} \times WI_{l,t} \qquad (14)$$

The total water usage per epoch $W_{tot,t}$ can be estimated by combining the individual water usages from the variety of sources discussed across all datacenters $L$:

$$W_{tot,t} = \sum_{l}^{L}(W_{E,l,t} + W_{B,l,t} + W_{Grid,l,t}) \quad (15)$$

### 3.4 Carbon Model

We model the carbon emissions that are present from both the use of electricity and water providing a holistic consideration of the carbon emissions of a datacenter. One way to measure the difference between energy usage and carbon emissions is to track the carbon intensity $CI_{l,t}$ of the underlying energy sources over the epoch $t$. $C_{Grid,l,t}$ is the amount of carbon emissions that come from the carbon intensity of the grid at datacenter location $l$:

$$C_{Grid,l,t} = CI_{l,t} \times E_{tot,t} \quad (16)$$

The generation of potable water and the processing of wastewater also makes use of energy that generates carbon emissions. The amount of energy that is utilized by the treatment facilities has been quantified in prior work and can be estimated by $EI_{pot,t}$ and $EI_{waste,t}$ [26]. Therefore, the carbon emissions during epoch $t$ directly associated with water usage are:

$$C_{W,l,t} = [(W_{B,l,t} + W_{E,l,t}) \times EI_{pot,t} + W_{G,l,t} \times EI_{waste,t}] \times CI_{l,t} \quad (17)$$

By combining these two sources of carbon emissions, the total emissions per time epoch $t$ can be found and the environmental impact over any epoch can be determined as:

$$C_{tot,t} = \sum_{l}^{L}(C_{Grid,l,t} + C_{W,l,t}) \quad (18)$$

## 4 Problem Formulation

We consider geo-distributed datacenters that are managed by a cloud service provider for supporting LLM inference workloads. A cloud management framework provides an LLM inference workload execution plan that includes workload assignment to each location and scheduling within the location. The goal of the framework is to co-optimize four objectives: TTFT, carbon emissions, water usage, and energy costs. We assume that workloads originate off-site from the datacenters. We consider the associated transfer time and costs based on assigned datacenter location to workloads. Once workloads are assigned to a location by the framework, a local datacenter scheduling policy is used to schedule local workloads. The local scheduling policy is designed to be fast and fair, based on a round-robin scheduling algorithm that is extended from [27]. It prioritizes workload scheduling based on their arrival time and determines processing destinations in a rational order.

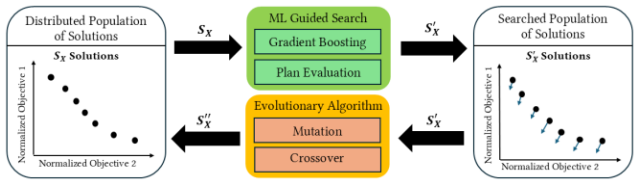

**Fig. 2: Overview of *SLIT* framework**

## 5 SLIT Framework Overview

Our proposed *SLIT* framework integrates a metaheuristic approach that utilizes machine learning (ML) assisted local search in combination with an evolutionary algorithm (EA) as shown in Fig. 2. The framework also includes a workload predictor to assist in ensuring that the scheduling plan accurately reflects the incoming workload. The pseudocode of our metaheuristic is shown in Fig. 3 and its line numbers are used in the following sections when discussing the components of the framework.

---
**Algorithm 1: *SLIT* Framework**

**Input:** $gen$, $iter_{early}$, $GradBoost$, $predictor$, $cluster$, $freq$
**Output:** $S$ scheduling plans up to $X$ final designs
**Initialization:**
$S_{init} = \{s_1 ..., s_X\}, Y_{traj} = \emptyset, S_{child} = \emptyset$
**Arrival Predictor:**
1. $predictor = best\_fit(predictor_{set})$
2. $prediction \leftarrow predictor$
**ML Guided Search:**
3. **for** $iteration = 0$ to $gen$ **do**
4.     $S_{start} \leftarrow ML(GradBoost, S_{init}, cluster, prediction)$
5.     **for** $s$ in $S_{start}$ **do**
6.         $(s_{new}, Y_{traj}) \leftarrow search(s, step)$
7.         $Y_{train} \leftarrow Y_{train} \cup Y_{traj}$
8.         $S \leftarrow update\_population(S, s_{new})$
9.     **end for**
10.   **if** $iter$ is divisible by $freq$
11.     $GradBoost \leftarrow train(Y_{train}), Y_{train} = \emptyset$
**Evolutionary Algorithm:**
12.   **for** $parents$ in $S$ **do**
13.     $s_{parent,1}, s_{parent,2} \leftarrow random\_select(S, 1)$
14.     $s_{child} \leftarrow cross\_over(s_{parent,1}, s_{parent,2})$
15.     $s_{child} \leftarrow s_{child} \cup mutate(s_{child})$
16.     $S_{child} \leftarrow update\_population(S_{child}, s_{child})$
17.     **for** $s_{child}$ in $S_{child}$ **do**
18.         $S \leftarrow update\_population(S, s_{child})$
19.     **end for**
20.   **end for**
21. **end for**
22. **If** $prediction \neq observed$ **do**
23.     $S_{missed} = S_{default}$, utilize scheduled default plan for missed requests
24. **return** $S$

---

**Fig. 3: Pseudocode for the *SLIT* framework**

### 5.1 Workload Predictor

Due to the high variability of the LLM workload intensity (as observed in Fig. 1), the use of a workload predictor is important. A workload predictor allows our framework to have an accurate workload to schedule for the upcoming epoch. Our $predictor$ is based on the regression-based predictor described in [28]. A set of linear regression models are put into $predict\_set$ (line 1) that are trained on the varying number of requests per epoch. To prevent overfitting onto the most recent epoch, the models are trained incrementally with the use of ML that selects the best model within the set $best\_fit$.

### 5.2 Machine Learning (ML) Guided Local Search

Local search algorithms are an effective method to find optimal solutions in a problem space. However, naïve approaches such as random local search are poor at quickly finding the optimal solution. These problems of long search times can be mitigated through the addition of ML guided search. For our framework, gradient boosting was used as the ML model which is an ensemble approach that utilizes multiple weak decision trees in its decision-making approach [29]. Gradient boosting is a powerful ML technique based on boosting in a functional space, where the target is pseudo-residuals instead of residuals as in traditional boosting. Its constituent ensemble of weak prediction models make very few assumptions about the data. We found this model to work well for our problem. The framework begins with a partially random initial population $S_{init}$ to begin searching from. This population includes two extreme scheduling plans (i.e. evenly distributed and only scheduled to one location) to promote the framework to search within a broader area of the solution space. Each scheduling plan $s$ within our population $S$ is searched (line 6), and the new point $s_{new}$ and search trajectory $Y_{traj}$ are generated. Each new scheduling plan $s_{new}$ is evaluated during the *update_population* step and only dominant plans are included in the updated population $S$ (line 8) The search trajectories $Y_{train}$ generated by searching are used to train a gradient boosting ML model (line 11). To prevent new search trajectories from undoing our previous training we limit training to occur at a set $freq$.

### 5.3 Evolutionary Algorithm

One disadvantage of relying only on the ML-guided search is the potential of getting trapped within local optima. This would occur when the solution provided is dominant when compared to all the immediate neighboring spaces but is still dominated by some other solution point elsewhere within the solution space. To allow our framework to more effectively search the entire solution space we introduce an evolutionary algorithm (EA) to allow learned knowledge from the ML-guided search to flow into unsearched regions. As all points within the population $S$ have been searched previously the parents $s_{parent,1}$ and $s_{parent,2}$ are randomly selected from the population (line 13). The two parents are then subject to being crossed over to share traits gained through the search process previously in a crossover step (line 14). This process generates a new scheduling plan $s_{child}$ which is further mutated through the random modification of the scheduling plan (line 15). This introduces new unseen plans that have not been evaluated before, allowing for the ML guided local search to explore new areas of the problem space. The children are then evaluated with only the dominant solutions being added and updating the population $S$ (line 18).

## 6 Experiments

We compare our proposed *SLIT* framework with two state-of-the-art LLM inference scheduling approaches: a mixed integer linear programming based framework (*Helix*) [16], and a queue based algorithm to schedule incoming requests (*Splitwise*) [17]. We consider six different types of nodes that are present across all the datacenters with between 2-8 GPUs of a homogenous type (i.e., NVIDIA A100 or H100). Each datacenter location has 1000 compute nodes available with an even amount of each type of node. Containers are launched with LLM models and handled using a serverless infrastructure. We consider 12 datacenters that are distributed globally across the regions of East Asia, Oceania, North America, and Western Europe. LLM requests can originate in any region. We developed and validated a Python-based simulator that integrates the models described in Section 3, along with support for configurable inputs and metric tracking.

We select and showcase five solutions from the final Pareto solution set generated by *SLIT*: the best single objective solutions (referred to as *SLIT-Carbon, SLIT-TTFT, SLIT-Water,* and *SLIT-Cost*) and a solution that balances the normalized values across all objectives (*SLIT-Balance*). Our experimental setup spans a 24-hour interval, with 0.5× the delay between requests, 3× the token count, and 10× the number of requests found in the LLM trace from [19]. For all frameworks, we assume that the workload predictor predicts LLM request arrival for the next epoch (15 minutes in the future). Thus, the maximum runtime of all frameworks is capped to 15 minutes, to provide real-time decisions.

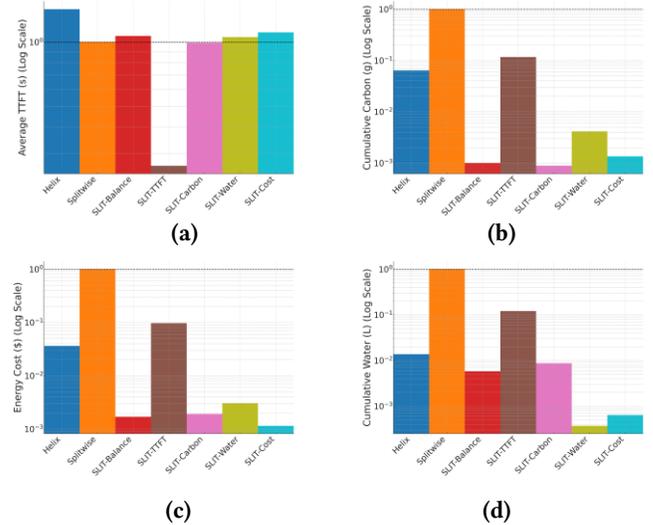

Fig. 4: Comparison of the normalized (a) TTFT, (b) carbon emission, (c) energy cost, and (d) water usage, across LLM inference scheduling frameworks. All plots are normalized to the results from the *Splitwise* framework.

Our main experimental results are presented in Fig. 4 and demonstrate how our *SLIT* framework with multiple scheduling solutions can match different needs of a cloud datacenter manager. Our single metric optimized solutions significantly outperform the *Helix* [16] and *Splitwise* [17] frameworks, with *SLIT-Carbon* reducing emissions by 98% vs *Helix* and 99% vs *Splitwise SLIT-Water* reducing water usage by 97% vs *Helix* and 99% vs *Splitwise*, *SLIT-TTFT* reducing TTFT by 81% vs *Helix* and 73% vs *Splitwise*, and *SLIT-Cost* reducing energy costs by 96% vs *Helix* and 99% vs *Splitwise*, respectively. However, all these single objective frameworks sacrifice other metrics to achieve these results. We can observe that if a balanced solution is desired, *SLIT-Balance* provides a balanced solution that outperforms *Helix* [16] in all metrics. When compared to *Splitwise* [17] it is able to provide lower carbon emissions, energy costs, and water usage, while still providing a competitive TTFT performance. In summary, we can observe that our *SLIT* framework is able to provide both great single-objective optimized solutions and balanced solutions across all the sustainability, cost, and performance objectives.

Fig. 5 shows a time-domain analysis of how solutions generated by *Helix* [16] and *Splitwise* [17] compare with the *SLIT-*

*Balance* solution. *Splitwise* and *SLIT-Balance* maintain very similar TTFTs throughout, but *SLIT-Balance* can achieve this without the high carbon emissions, water usage, and energy costs observed for *Splitwise*. In contrast, we observe that *Helix* is worse in all metrics when compared to *SLIT-Balance* on a per epoch basis.

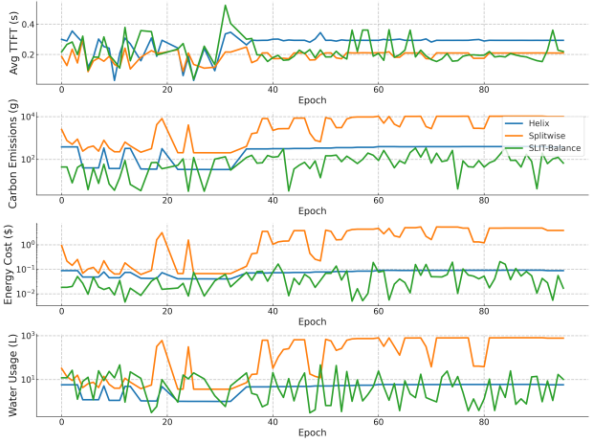

**Fig. 5: Per-epoch metric comparison for LLM scheduling across the *Helix, Splitwise*, and *SLIT* frameworks.**

## 7 Conclusion

In this work we presented our novel framework called *SLIT* to address the problem of LLM inference request scheduling across geo-distributed datacenters to co-optimize the time-to-first-token (TTFT), carbon emissions, water usage, and energy costs. We developed comprehensive models of LLM execution, inter-datacenter latency, energy consumption, water use, carbon emissions, and energy costs in the context of server nodes with GPUs utilized for LLM inference. From our experiments we observed that *SLIT* was able to produce diverse high-quality solutions that outperformed state-of-the-art frameworks.

### ACKNOWLEDGMENTS

This research was supported by HPE and grants from the National Science Foundation (CCF-2324514, CNS2132385)